\documentclass[a4paper,fleqn,usenatbib]{mnras}

\usepackage[T1]{fontenc}
\usepackage{ae,aecompl}

\usepackage{graphicx}	
\usepackage{amsmath}	
\usepackage{amssymb}	
\usepackage{booktabs}
\usepackage[utf8]{inputenc}

\newcommand{\changed}[1]{{#1}}

\title[Tomographic laser guide star tip-tilt determination]{A tomographic algorithm to determine tip-tilt information from laser guide stars}

\author[A.~P.~Reeves, T.\@J.Morris, R. M. Myers, N. A. Bharmal and J. Osborn]{A. P. Reeves,$^{1}$\thanks{E-mail:
a.p.reeves@durham.ac.uk (APR);} T. J. Morris,$^{1}$ R. M. Myers,$^{1}$ N. A. Bharmal$^{1}$ and J. Osborn$^{1}$.\\
\\
$^{1}$Centre for Advanced Instrumentation, Durham University, Durham, DH1 3LE, UK\\
}

\date{Accepted XXX. Received YYY; in original form ZZZ}

\pubyear{2015}

\begin{document}
\label{firstpage}
\pagerange{\pageref{firstpage}--\pageref{lastpage}}
\maketitle

\begin{abstract}
    Laser Guide Stars (LGS) have greatly increased the sky-coverage of Adaptive Optics (AO) systems. Due to the up-link turbulence experienced by LGSs, a Natural Guide Star (NGS) is still required, preventing full sky-coverage. We present a method of obtaining \changed{partial} tip-tilt information from LGSs alone in multi-LGS tomographic LGS AO systems. The method of LGS up-link tip-tilt determination is derived using a geometric approach, then an alteration to the Learn and Apply algorithm for tomographic AO is made to accommodate up-link tip-tilt. Simulation results are presented, verifying that the technique shows good performance in correcting high altitude tip-tilt, but not that from low altitudes. We suggest that the method is combined with \changed{multiple far off-axis tip-tilt NGSs} to provide gains in performance and sky-coverage over current tomographic AO systems.
\end{abstract}



	\section{Introduction} 
	The use of Laser Guide Stars (LGSs) in Adaptive Optics (AO) has greatly increased the area of the sky available for correction, from $\mathrm{\approx10\%}$ up to $\mathrm{\approx 85\%}$ \citep{ellerbroek1998}. In turn this has led to a vast increase in the number of astronomical science targets which can be observed using AO. The laser experiences turbulence whilst traveling upwards to form an artificial guide star, so its position will move in the sky. It is thought that this effect renders all `tip-tilt' information gained from LGS Wave-Front Sensors (WFS) useless, as it is a function of LGS `up-link' movement and the desired `down-link' tip-tilt, which have previously been considered to be entangled irretrievably. It has even been suggested that the tip-tilt the laser acquires on the up-link is the reciprocal of the global tip-tilt on the down-link path, hence little tip-tilt will be observed on the WFS at all \changed{\citep{pilkington1987}}. To correctly obtain the science path tip-tilt, a Natural Guide Star (NGS) must still be used \changed{\citep{rigaut1992}}. As a tip-tilt WFS requires relatively few photons and the anisoplanatic patch size is large for tip-tilt modes, the requirements on a NGS are much lessened \citep{wilson1996}. Nonetheless, requiring a tip-tilt NGS still limits the sky-coverage of an LGS AO system.
	
	Tomographic LGS configurations, such as Laser Tomographic AO (LTAO)\citep{stroebele2006}, Multi-Object AO (MOAO) \citep{conan2010,gendron2011}  and Multi-Conjugate AO (MCAO) \citep{marchetti2003, rigaut2012}, are coming online. These AO configurations use information from a number of LGSs, off-axis from the science target, to estimate the science path turbulence. Such systems overcome the so called cone-effect where the LGS samples a cone of turbulence in the science path rather than the full cylinder of turbulence seen by light from the science target \citep{foy1985}. They can also achieve a large corrected field of view in the case of MCAO, or a large `field of regard' in the case of MOAO. Tomographic LGS systems still require a NGS to estimate tip-tilt modes, limiting their potential sky coverage. Suitable NGS are notoriously absent from much of the sky around the galactic poles, where many scientifically interesting targets exist \citep{ellerbroek1998}. 
    
    Methods to obtain all correction information from LGSs alone have been proposed. \changed{Some have not yet been implemented due to the requirement of complex laser schemes and/or auxiliary beam viewing telescopes \citep{ragazzoni1995, belen'kii1995, foy1995, ragazzoni1996a, ragazzoni1997, belenkii1997}}. \changed{\citet{ragazzoni1996b} discusses the use of the temporal delay between the launch time of a laser pulse and the time it is received by the telescope to estimate up-link tip-tilt. This technique requires the use of the full telescope aperture for laser launch, which implies some problems with scattered light and fluorescence.} More recently, \citet{basden2014} has proposed an LGS assisted lucky imaging system, which could provide full sky AO coverage but entails discarding some science flux and would not be suitable for spectroscopy. \citet{davies2008} explored the potential usage of LGS AO with no tip-tilt signal, allowing 100\% sky-coverage. It was found that for some applications, a dedicated NGS tip-tilt star was not required and a telescopes fast guiding system was adequate. 
    
    \changed{In this paper we propose a method to retrieve partial tip-tilt information from a number of LGSs in existing or currently proposed tomographic AO systems from only the systems wavefront sensor measurements. This is similar to that proposed by \citet{ragazzoni1998}, though we do not required the knowledge of the exact sky position of one of the LGS. We aim to improve AO performance across the whole sky over AO performed with no tip-tilt NGS, potentially relaxing the requirement for, or for some applications eliminating the need for, a NGS.}
    
    If the tip-tilt modes measured across the full-aperture are the same as that across the beam launch aperture, it is clear that the tip-tilt signal would indeed be irretrievable from LGS WFSs, as they would be reciprocal and little atmospheric tip-tilt would be observed from a LGS WFS at all. In § \ref{sec:TTCorr}, we show that the tip-tilt modes across the beam launch telescope are uncorrelated with those over the whole aperture, opening the possibility of LGS up-link tip-tilt determination. The algorithm for up-link tip-tilt determination is derived in § \ref{sec:lgsTT} and an adaptation to the LA  algorithm proposed by \citet{Vidal2010} for MOAO is suggested as a practical method for its use. Results from simulation verifying the technique are presented in § \ref{sec:simRes}. Finally, in § \ref{sec:discussion} we discuss the practical uses of an LGS up-link retrieving AO system, and how it may provide increases in LTAO sky-coverage and performance by combination with ground layer tip-tilt correction.

	\section{Correlation of Tip-Tilt Between Telescope and Beam-Launch Apertures}
	\label{sec:TTCorr}

	If the global tip-tilt across the telescope aperture is identical to that over the beam launch telescope, any tip-tilt encountered by the LGS up-link path will have an equal but opposite effect on the return path \changed{(though with a slight change due to temporal delay \citep{ragazzoni1996b})}. Consequently \changed{little} tip-tilt will be observed on the LGS WFS and the tip-tilt component of the science path can not be determined by that WFS. This is referred to as tip-tilt `reciprocity' and is the case if the laser is launched from the full aperture of the telescope. All current facility LGS AO systems use a separate Laser beam Launch Telescope (LLT). On these telescopes and those planned for the future, \changed{$\mathit{D_{LLT}} << \mathit{D}$}, where $D_{LLT}$ denotes the diameter of the LLT and $D$ is the size of the telescope aperture.
	
	 Determination of LGS up-link tip-tilt can only be possible if the up-link and down-link tip-tilt components are uncorrelated or it will not be fully observed by the WFS. The covariance between two concentric Zernike modes of different radii in Kolmogorov turbulence is shown in equation (\ref{eq:wj_ZCov}) \citep{wilson1996},
	\begin{multline}\label{eq:wj_ZCov}
		\mathrm{C} = 0.0145786 \it{e}^{\frac{1}{2} i \pi  (n-p)} \sqrt{(n+1) (p+1)} \left(\frac{R}{\mathrm{r_0}}\right)^{5/3}  \\ \times \int_0^\infty dk \frac{J_{n+1}(2  \pi k ) J_{p+1}(2 \pi  \gamma k)}{\gamma  k^{14/3}},
	\end{multline}
where $\gamma$ represents the fractional size relationship between the two apertures, $\mathit{n}$ and $\mathit{p}$ are the radial orders of the two Zernike polynomials, $R$ is the radius of the telescope, $J_{n+1}$ and $J_{p+1}$ are Bessel functions of the first kind, $k$ is the wave number of the light and $\mathit{r_0}$ is the atmospheric Fried parameter \citep{fried1966}.
		
\begin{figure}
	\includegraphics[width=0.5\textwidth]{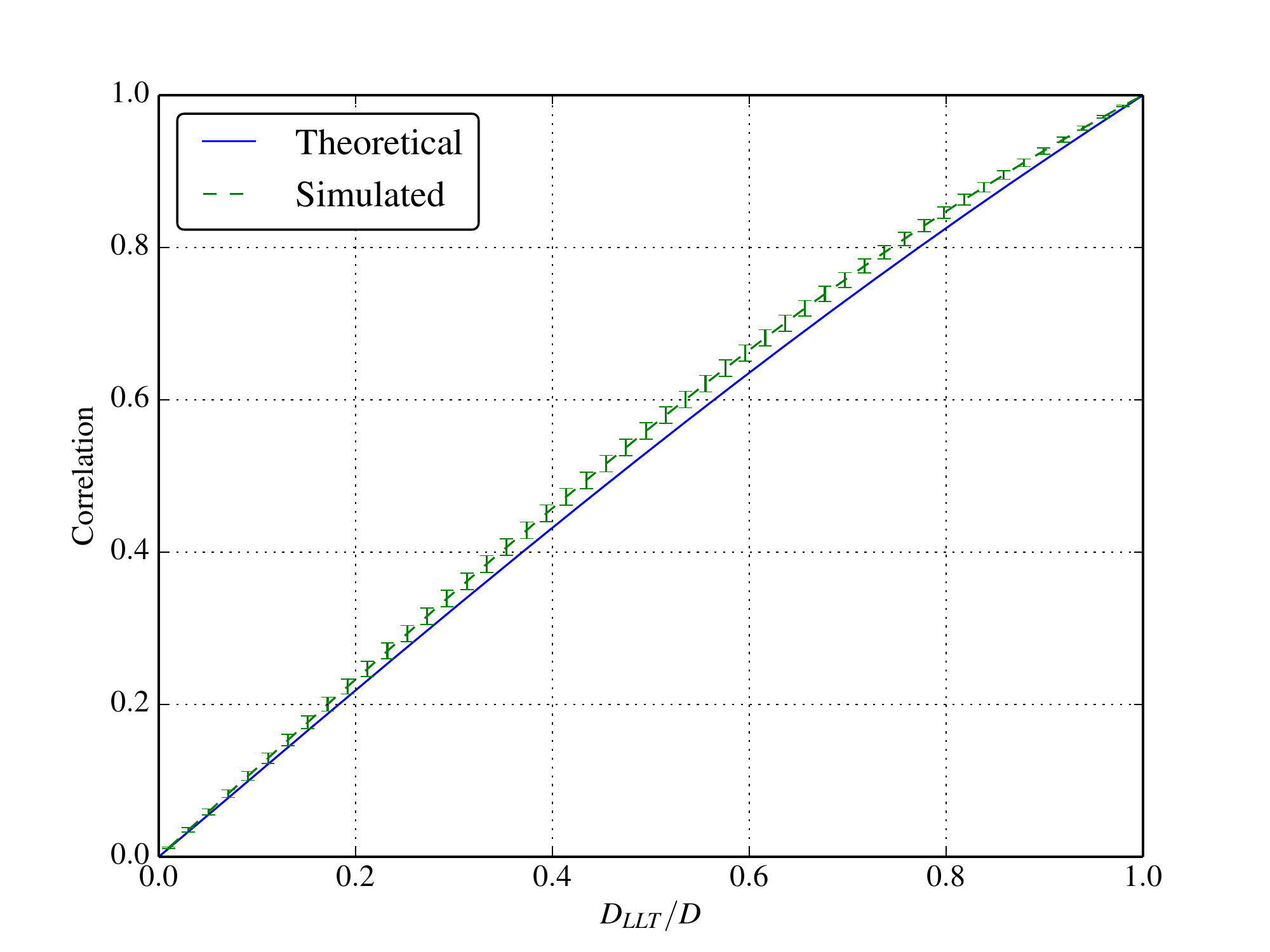}
	\caption{Theoretical and simulated correlation of phase perturbations in Kolmogorov turbulence between concentric tip-tilt modes as a function LLT diameter, $D_{LLT}$, as a fraction of the telescope diameter, $D$.}
	\label{fig:TTCorPlot}
\end{figure}

The covariance between concentric tip-tilt modes of different radii are plotted in Figure \ref{fig:TTCorPlot} (where $\mathit{n,p} = 1$). A plot of the correlation of tip-tilt modes in ten thousand simulated random Kolmogorov phase screens is also plotted. It is evident that the correlation of tip-tilt modes between small and large apertures in the regime where $\mathit{D_{LLT}}/\mathit{D} < 0.1$ is less than $0.1$. \changed{This is true for both the theoretical expression and the simulated phase screens, which match closely in such a regime}. This result means that tip-tilt modes will not be reciprocal and will be visible on an LGS WFS. Observed tip-tilt will be some function of the turbulence encountered by the laser as it propagates up to form an artificial guide star and the global tip-tilt across the telescope aperture as it propagates back. \changed{It should be noted that the tip-tilt observed by the laser will be larger than that seen by the full aperture due to greater spatial averaging on larger scales.}

\changed{Despite being uncorrelated, it is possible that by chance there is a similar component of tip-tilt across both the launch and telescope apertures, in which case it will not be observed on a LGS WFS. If multiple LGSs are used, then the chance of the same component of tip-tilt being present across all launch paths is significantly reduced.}

\section{Tomographic LGS Tip-Tilt Determination}
	\label{sec:lgsTT}
	
	\subsection{Retrieving down-link turbulence induced slopes}
	\label{sec:lgsTT_1}
	As demonstrated in the previous section, the measurement from a LGS WFS is a function of the atmospheric turbulence the laser propagates through on the way up to form a guide star and the turbulence the return light propagates through as it travels back down to the telescope. For AO correction of an astronomical science target the two components must be separated and only the latter is required. If using a Shack-Hartmann or Pyramid WFS, WFS measurements will be in the form of slopes representing the gradients of the measured phase within any given sub-aperture. The use of such a gradient measuring WFS is assumed in the following derivations. We can express slopes measured by an LGS WFS as the sum of the laser up-link induced slopes and the down-link turbulence induced slopes,
    \begin{equation}\label{eq:def_s}
		\mathbf{\tilde{s}} = \mathbf{\tilde{s}_l}+\mathbf{\tilde{s}_t},
	\end{equation}
    
where $\mathbf{\tilde{s}}$ is a vector representing the slopes measured on a WFS, $\mathbf{\tilde{s}_l}$ is a vector representing the slopes caused by LGS up-link turbulence and $\mathbf{\tilde{s}_t}$ is a vector representing the slopes caused by down-link turbulence. For AO correction of a natural astronomical science target we must obtain $\mathbf{\tilde{s}_t}$. Note that LGS up-link turbulence results exclusively in tip-tilt modes being observed on the WFS and no higher order modes, so $\mathbf{\tilde{s}_l}$ will be homogeneous in the $x$ and $y$ directions. For an AO system with a single LGS and no external reference, determining $\mathbf{\tilde{s}_t}$ is not possible as there is not enough information to determine $\mathbf{\tilde{s}_l}$. In a tomographic system, there is more information about the turbulence sampled by the LGSs on the up-link, and $\mathbf{\tilde{s}_t}$ can be computed. 
	
	\begin{figure}
		\includegraphics[width=0.5\textwidth]{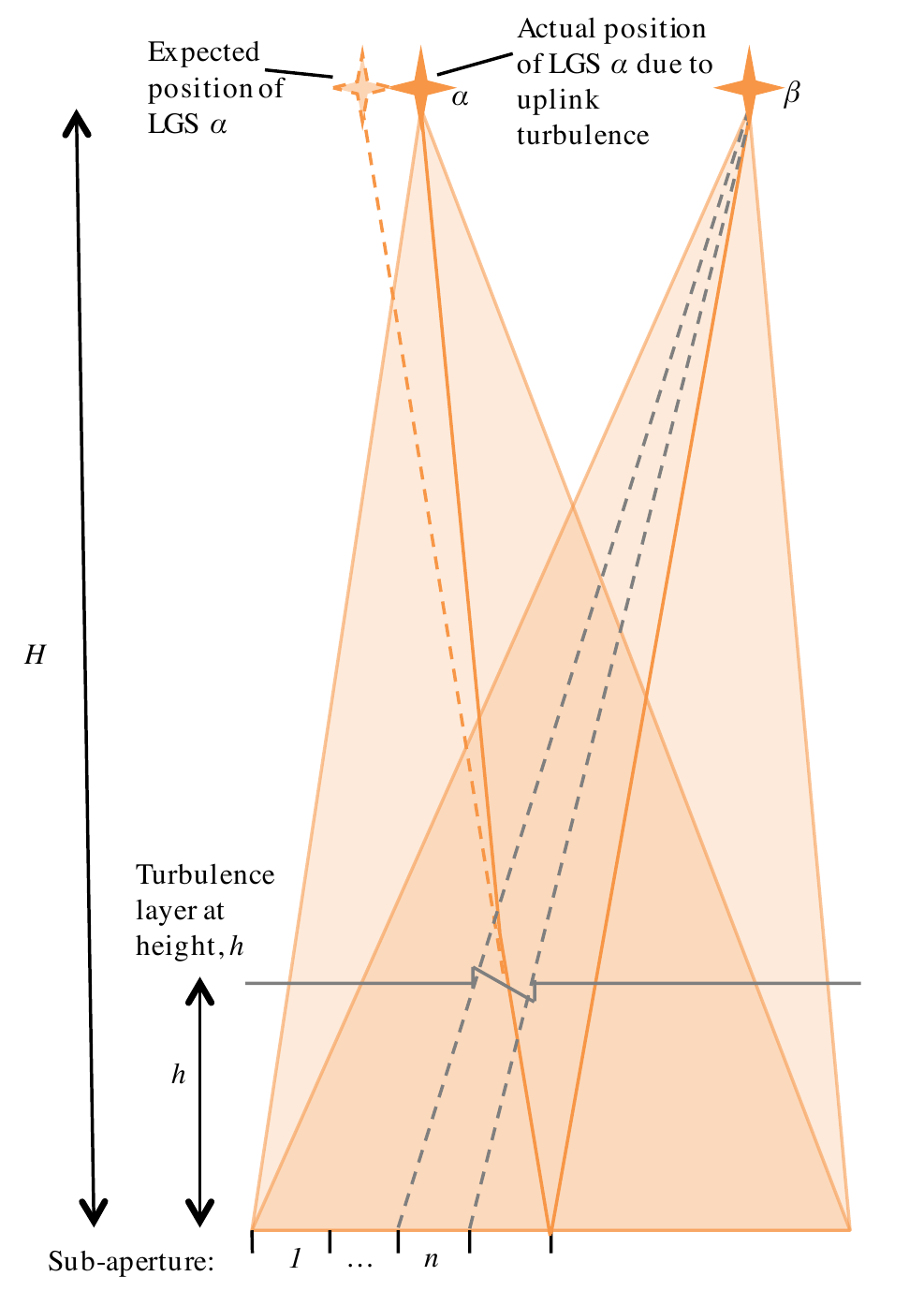}
		\caption{The geometry of the LGS system under consideration. One turbulent layer is shown, which features only a tilt at the point that LGS $\alpha$ overlaps with the field of view of sub-aperture $n$ on WFS $\beta$. }
		\label{fig:overlapping_LGS}
		\end{figure}
For the remainder of this section we consider a trivial 2-dimensional, tomographic, two LGS AO geometry, where both LGSs are centre-launched. The following approach can be scaled to many centre launched LGSs, though the mathematics quickly becomes cumbersome with more than three. The LGSs are labeled LGS $\alpha$ and LGS $\beta$ and the observing WFSs as WFS $\mathrm{\alpha}$ and WFS $\mathrm{\beta}$. Slopes measured on WFSs are denoted as $\mathbf{\tilde{s}_\alpha}$ and $\mathbf{\tilde{s}_\beta}$ respectively. This geometry is illustrated in Figure \ref{fig:overlapping_LGS}.

WFS $\beta$ observes the area of turbulence which causes the up-link tilt on WFS $\alpha$, hence we postulate that there is a transform $\mathbf{\hat{T}_{\alpha \beta}}$ which relates the down-link turbulence induced slopes, $\mathbf{\tilde{s}_{\beta  t}}$, to up-link induced tip-tilt measured on WFS $\alpha$, $\mathbf{\tilde{s}_{\alpha l}}$,
	\begin{equation}\label{eq:Tdef1}
		\mathbf{\tilde{s}_{\alpha l}} = \mathbf{\hat{T}_{\alpha\beta}} \mathbf{\tilde{s}_{\beta t}}.
	\end{equation}

    We initially consider the simple situation illustrated in Figure \ref{fig:overlapping_LGS}, where a single turbulent layer at a height $h$, which features only a tilt in the section where LGS $\mathrm{\alpha}$ overlaps with the field of view (FOV) of sub-aperture $n$ on the WFS observing LGS $\mathrm{\alpha}$. In this case it is clear that such a transform, $\mathbf{\hat{T}_{\alpha \beta}}$, exists and can be trivially computed as WFS $\beta$ is unaffected by up-link turbulence so $\mathbf{\tilde{s}_{\beta l}} = 0$, hence $\mathbf{\tilde{s}_{\alpha l}} = \mathbf{\hat{T}_{\alpha\beta}} \mathbf{\tilde{s}_{\beta}}$. In general however $\mathbf{\tilde{s}_{\beta t}}$ will not be known, as LGS $\beta$ will also experience up-link tip-tilt. For this general case, 
	\begin{align}\label{eq:sub1}
		\mathbf{\hat{T}_{\alpha \beta }} \mathbf{\tilde{s}_{\beta}} &= \mathbf{\hat{T}_{\alpha \beta }} (\mathbf{\tilde{s}_{\beta t} + \mathbf{\tilde{s}_{\beta l}}}) \nonumber \\
		\mathbf{\hat{T}_{\alpha \beta }} \mathbf{\tilde{s}_{\beta}} &= \mathbf{\tilde{s}_{\alpha l}} + \mathbf{\hat{T}_{\alpha \beta }} \mathbf{\tilde{s}_{\beta l}}
		\end{align}
and
	\begin{equation}\label{eq:sub2}
		\mathbf{\hat{T}_{ \beta \alpha }} \mathbf{\tilde{s}_{\alpha}} = \mathbf{\tilde{s}_{\beta l}} + \mathbf{\hat{T}_{ \beta \alpha}} \mathrm{\tilde{s}_{\alpha l}}.
	\end{equation}
    
    We now have 2 equations in (\ref{eq:sub1}) and (\ref{eq:sub2}), to solve for 2 unknowns, $\mathbf{\tilde{s}_{\alpha l}}$ and $\mathbf{\tilde{s}_{\beta l}}$. Re-arranging equation (\ref{eq:sub2}),
\begin{equation}
		\mathbf{\tilde{s}_{\alpha l}} = \mathbf{\hat{T}_{\beta \alpha }^{-1}} ( \mathbf{\hat{T}_{ \beta \alpha}} \mathbf{\tilde{s}_{\alpha}} - \mathrm{\tilde{s}_{\beta l}} )
\end{equation}
and substituting into equation (\ref{eq:sub1}),
\begin{align}
		\mathbf{\hat{T}_{\alpha \beta }} \mathbf{\tilde{s}_{\beta}} &= \mathbf{\hat{T}_{\beta \alpha}^{-1}} (\mathbf{\hat{T}_{ \beta \alpha}} \mathbf{\tilde{s}_{\alpha}} - \mathrm{\tilde{s}_{\beta l}} ) + \mathbf{\hat{T}_{\alpha \beta}} \mathrm{\tilde{s}_{\beta l}} \nonumber \\
		&=  \mathbf{\tilde{s}_{\alpha}} + (\mathbf{\hat{T}_{\alpha \beta}} - \mathbf{\hat{T}_{ \beta \alpha}}^{-1}) \mathrm{\tilde{s}_{\beta l}}.
\end{align}
Finally, re-arranging for $\mathbf{\tilde{s}_{\beta l}}$, 
\begin{equation}\label{eq:s_bl}
		\mathbf{\tilde{s}_{\beta l}} = ( \mathbf{\hat{T}_{\alpha \beta}} - \mathbf{\hat{T}_{\beta \alpha}}^{-1} )^{-1} ( \mathbf{\hat{T}_{\alpha \beta}} \mathbf{\tilde{s}_{\beta}} - \mathbf{\tilde{s}_{\alpha}} )
\end{equation}
and similarly for $\mathbf{\tilde{s}_{\alpha l}}$
\begin{equation}
	\mathbf{\tilde{s}_{\alpha l}} = (\mathbf{\hat{T}_{\beta \alpha}} - \mathbf{\hat{T}_{\alpha \beta}}^{-1} )^{-1} (\mathbf{\hat{T}_{\beta \alpha}} \mathbf{\tilde{s}_{\alpha}} - \mathbf{\tilde{s}_{\beta}}).
	\end{equation}

 $\mathbf{\tilde{s}_{\alpha}}$ and $\mathbf{\tilde{s}_{\beta}}$ are the WFS measurements and the $\mathbf{\hat{T}}$ transforms can be obtained by considering the geometry of the system i.e., where sub-apertures from a WFS observe the up-link path of the other laser(s). 
 It is now possible to calculate the turbulence induced slopes, as $\mathbf{\tilde{s}_t} = \mathbf{\tilde{s}} - \mathbf{\tilde{s}_l}$. These are the slopes which would have been measured from a guide star with no up-link tip-tilt effects, and can be used to perform the AO reconstruction without the requirement of an NGS for tip-tilt measurement. The above analysis can be performed for more complex LGS AO systems with many LGSs in other geometries.
 
In general there will be more than one discrete turbulent layer in the atmosphere, hence the measurement of a particular element in $\mathrm{\tilde{s}_{\beta t}}$ which overlaps with LGS $\alpha$ will not just represent the turbulence at height $h$, but will be the sum of measurements from all turbulent layers. This represents some noise in the measurement of $\mathrm{\tilde{s}_{\alpha l}}$. The noise is mitigated by increasing the number of LGSs, such that other layers from non-overlapping heights average to zero, leaving only the common measurement of the slope at the point LGS $\alpha$ overlaps with the layer at altitude $h$. 
 
For the centre launched case, the slopes due to down-link turbulence, $\mathbf{s_t}$, cannot be determined for a turbulent layer at the ground. For a layer at this height, $\mathbf{\tilde{s}_{\alpha t}}=\mathbf{\tilde{s}_{\beta t}}$, $\mathbf{\tilde{s}_{\alpha l}}=\mathbf{\tilde{s}_{\beta l}}$, $\mathbf{\tilde{s}_{\alpha}}=\mathbf{\tilde{s}_{\beta}}$, and there is no-longer more than one independent equation from which to determine $\mathbf{\tilde{s}_{\alpha l}}$ and $\mathbf{\tilde{s}_{\beta l}}$. \changed{Further to this issue, if the lasers are centre launched, then the beam paths are likely to be obscured by the ``shadow'' of the secondary obscuration, the turbulence that the laser passes near the ground layer is not measured.}

\changed{An AO system that launches LGSs from different points within the telescope aperture could potentially overcome these limitations. In this case $\mathbf{\tilde{s}_{\alpha l}} \neq \mathbf{\tilde{s}_{\beta l}}$ at the ground layer, so both can be determined. Depending on the laser launch scheme, it is also possible that the low layer turbulence in the beam path could be observed. Such as scheme does however entail other difficulties, such as scattering from launching the laser directly off the primary mirror.}

A system with LGSs launched from outside the telescope aperture (side launched) is unlikely to be suitable for this method of LGS up-link tip-tilt correction as a LGS's launch path is not observed by other LGS WFSs. It is possible that outer WFS sub-apertures could be used as they may correlate strongly with the launch path turbulence, though this is outside the scope of this work.

\subsection{Obtaining LGS up-link transforms}
	
	The LGS up-link matrices describing the response of LGS motion to WFS measurements are defined in equation (\ref{eq:Tdef1}), they relate down-link turbulence measurements of a WFS to the predicted up-link path of another LGS. They can be calculated by considering the geometry illustrated in Figure \ref{fig:overlapping_LGS} and the effect of a layer at a height $h$ with a small region of turbulence where the FOV of sub-aperture $n$ overlaps with LGS beam $\alpha$, at height $H$. 
	
	For a given sub-aperture, $s_{\alpha l}$ is the slope measured due to up-link tip-tilt on WFS $\alpha$. It is related to the slope measured on a corresponding sub-aperture on WFS $\beta$ which views LGS $\alpha$ at height $h$, $s_{\beta t}$.  It is shown in Appendix \ref{append:Interaction} that
	\begin{equation}
		s_{\alpha l} = \frac{\lambda}{2 \pi} \frac{H-h}{H} s_{\beta t}.
	\end{equation}

	The system has only a finite vertical resolution (defined by the number of sub-apertures and LGS asterism separation) to correct for turbulent layers. Sub-aperture $n$ will observe the integrated turbulence between where the LGS enters its FOV and where is exits its FOV. Different groups of sub-apertures on a WFS will correspond to different turbulent layer heights for which they can predict up-link tip-tilt on other LGS WFSs. Recalling that $\mathbf{s}_{\alpha l}$ is homogeneous in each direction, $\mathbf{\hat{T}_{\alpha \beta}}$ is 
	\begin{equation}
		\label{eq:lgs_int_mat}
	\mathbf{\hat{T}_{\alpha \beta}} = \frac{\lambda}{2\pi} 
	\begin{pmatrix}
	\frac{H-h_1}{H} & \frac{H-h_2}{H} & \cdots & \frac{H-h_N}{H} \\
	\frac{H-h_1}{H} & \frac{H-h_2}{H} & \cdots & \frac{H-h_N}{H} \\
	\vdots & \vdots & \ddots & \vdots \\
	\frac{H-h_1}{H} & \frac{H-h_2}{H} & \cdots & \frac{H-h_N}{H}
	\end{pmatrix},
	\end{equation}
where $h_n$ denotes the centre of the vertical height `bin' resolvable by the sub-aperture $n$. By considering the system geometry, including the launch angle of LGS $\alpha$ and $\beta$, $\theta_\alpha$ and $\theta_\beta$ respectively, we show in Appendix \ref{append:binHeight} that $h_n$ can be expressed as
	\begin{equation}
		h_n = \frac{H(\frac{D}{2} - (n+0.5)d)}{\frac{D}{2} - (n+0.5)d+H(\theta_{\alpha}+\theta_{\beta})}.
	\end{equation}
Again, this relationship can be extended for any number of LGS WFSs, where WFS $\beta$ observes the path of LGS $\alpha$.

The final step in creating an LGS up-link transform is to tailor the matrix to the required atmospheric turbulence vertical profile. Each column in the matrix shown in equation (\ref{eq:lgs_int_mat}) represents a vertical height bin resolvable by the tomographic LGS AO system. If a turbulence profile is known, then columns that represent heights where there is negligible turbulence can be set to zero. This step will reduce the noise contributed by `false layers' which could otherwise be detected, where random perturbations from real turbulent layers or noise could seem like turbulence at a height where no turbulence is present. Such a profile can be obtained from either the tomographic AO system itself or an external profiling instrument. 

\section{A Learn and Apply Approach}\label{sec:LA}

The geometric approach described in the previous sections to estimate and recover LGS tip-tilt modes is clearly highly idealised. It requires knowledge of the turbulence $\mathrm{C_n^2}$ vertical profile and that the calibration of the LGS WFSs and pointing of the LGSs is perfect. It would also not take into account our understanding of atmospheric turbulence statistics to improve correction. As correlation between adjacent sub-apertures can be significant \citep{wilson1996}, information from sub-apertures around those which view the up-link path of another LGS can be used to improve estimation of its up-link tip-tilt. \changed{This also allows an estimate to be made of ground layer tip-tilt, even for centre launched LGS AO systems, from sub-apertures surrounding the central obscuration. }

Learn and Apply (LA) is a method used in tomographic AO systems, such as MOAO and LTAO, for open-loop tomographic reconstruction which accounts for atmospheric turbulence statistics and the calibration of an AO system \citep{Vidal2010}. Instead of using a purely geometrical approach for LGS up-link tip-tilt determination, LA can be modified to implicitly account for up-link tip-tilt. LA is briefly described below.

If there is a linear relationship between off-axis WFS measurements, $\mathbf{\tilde{s}_{off}}$, and WFS measurements on-axis to the direction of a science target, $\mathbf{\tilde{s}_{on}}$, one can write
\begin{equation}
	\mathbf{\tilde{s}_{on}} = \mathbf{\hat{W}} . \mathbf{\tilde{s}_{off}}
\end{equation}
where $\mathbf{\hat{W}}$ is the tomographic reconstructor. If $\mathbf{\hat{W}}$ can be obtained, it can be used to calculate pseudo WFS measurements in the direction of a potential science target, which can then be used to calculate DM commands to provide correction in that direction. To estimate it, a large number of uncorrected WFS measurements of both on and off-axis slopes may be taken. In this case, the set of on-axis measurements are denoted as $\mathbf{\hat{M}_{on}}$ and the set of off-axis measurements as $\mathbf{\hat{M}_{off}}$. Vidal et al. show that a tomographic reconstructor for these specific measurements, $\mathbf{\hat{W'}}$, can be expressed as 
\begin{equation}
    \mathbf{\mathbf{\hat{W'}}} = (\mathbf{\hat{M}_{on} \hat{M}_{off}^{t}}) (\mathbf{\hat{M}_{off} \hat{M}_{off}^{t}})^{-1}.
\end{equation}

If the number of measurements taken to form  $\mathbf{\hat{M}_{on}}$ and $\mathbf{\hat{M}_{off}}$ approaches infinity, the tomographic reconstructor $\mathbf{\hat{W'}}$ approaches $\mathbf{\hat{W}}$, a general reconstructor which can reconstruct any set of slopes for the given guide star geometry and atmospheric turbulence profile. In this limit, the expressions $\mathbf{\hat{M}_{on} \hat{M}_{off}^{t}}$ and $\mathbf{\hat{M}_{off} \hat{M}_{off}^{t}}$ approach $\mathbf{\hat{C}_{OnOff}}$ and $\mathbf{\hat{C}_{OffOff}}$, the covariance matrices between on-axis and off-axis slopes, and off-axis and off-axis slopes respectively. The generalised tomographic reconstructor may now be expressed as
\begin{equation}
	\mathbf{\hat{W}} = \mathbf{\hat{C}_{OnOff}} \times \mathbf{\hat{C}_{OffOff}}^{-1}.
\end{equation}

If the profile of the atmosphere and system calibration is well known, both covariance matrices can be calculated purely analytically from statistical descriptions of turbulence. As this situation is not often the case, an alternative is to record some data from the system in open loop to create a `raw' covariance matrix which contains information regarding the atmospheric profile and AO system calibration. The `raw' covariance matrix cannot be used alone to create the tomographic reconstructor as it is not generalised and would also contain errors due to noise. It can though be used to act as a reference to fit an analytically generated covariance matrix which is generalised and not prone to noise effects. In this way LA creates a generalised tomographic reconstructor which accounts for AO system calibration and our statistical description of atmospheric turbulence. This process is termed the `learn' stage of the reconstructor.

Once both covariance matrices have been computed, the reconstructor, $\mathbf{\hat{W}}$, can be formed and `applied' to off-axis slopes to give an estimation of on-axis slopes. The LA algorithm has been tested successfully both in the laboratory and on-sky by the CANARY MOAO demonstrator \citep{gendron2011}.

We propose that the LA algorithm is also applicable for LGS tip-tilt determination, as it was demonstrated in § \ref{sec:lgsTT} that the required science direction slopes are a linear function of the off-axis LGS measurements. The advantages of using LA are many fold. LGS tip-tilt determination can account for system alignment and LGS pointing. The mathematics shown in § \ref{sec:lgsTT} does not have to be repeated for higher numbers of LGS, which quickly becomes cumbersome. The turbulence profile does not have to be externally measured to a very high vertical resolution. Finally and perhaps most importantly, the use of covariance matrices implicitly includes information about LGS up-link from sub-apertures near to those identified as geometrically observing a LGS beam.

To use LA, it must be altered to account for the fact that the tip-tilt signal from LGS WFSs is no longer removed. The analytical form of slope covariance matrices in this case must be derived. We consider the covariance between two WFS separated slope measurements with the definition given in equation (\ref{eq:def_s}),
\begin{align}\label{eq:covmat}
    \langle s_{\alpha} s_{\beta} \rangle &= \langle (s_{\alpha t} + s_{\alpha l})(s_{\beta t} + s_{\beta l}) \rangle \nonumber \\
    &= \langle s_{\alpha t} s_{\beta t} \rangle + \langle s_{\alpha t} s_{\beta l}\rangle  + \langle s_{\alpha l} s_{\beta t}  \rangle + \langle s_{\alpha l} s_{\beta l} \rangle.
\end{align}
Of these terms, the first is only a result of down-link turbulence and is the same as the covariance matrix which would be required in conventional Learn and Apply. This term can be calculated in a form similar to that which \citet{Vidal2010} use to obtain the covariance matrices between separated NGS WFS measurements with some modification to account for the cone effect associated with LGSs.

The second and third terms describe the relationship between the observed down-link turbulence and the tip and tilt observed by another WFSs due to the patch of turbulence that the lasers pass through on their up-link paths. They can be calculated again by considering the separation of each sub-aperture on the down-link with the launch path for each laser. As they are formed by a large tip or tilt from one WFS correlating with measurements from a single, or small number of, sub-aperture(s) from another WFS, it is expected that they will appear as a matrix of vertical and horizontal stripes.

The final term is the covariance between the up-link induced tip-tilt measurements. This value is dependent upon the separation between the two laser paths at an altitude layer and as it is a result of only tip and tilt, it is constant for each pair of WFSs. Assuming a centre launched case, this term will be large for low altitude layers, where the up-link laser paths overlap and small at high layers where the laser paths are separated. As it is constant, this value reduces the contrast of the the covariance matrices and so effectively make them less useful. Hence, it is again expected that this approach will work well at higher layers where this term is small, but less well for low layers where it will dominate.

The simulations performed in § \ref{sec:simRes} use a LA approach to predict LGS up-link tip-tilt. We do not yet attempt to derive the analytical form of the required covariance matrices. We instead rely on the fact that we can simulate a very large number of uncorrelated phase screens to create a large set of `learn' slopes to compute covariance matrices between off-axis and on-axis slopes. Deriving the analytical covariance matrices could improve on the performance we show and would be essential for use on-sky.

\section{SIMULATION RESULTS}
\label{sec:simRes}

\subsection{Simulation set-up}
In the following simulations we use the modified, tip-tilt retrieval LA based algorithm to perform LGS AO correction with no NGS tip-tilt WFS and compare these results to the currently used LTAO configuration, where tip-tilt information is removed from LGS slopes, and a low-order NGS used to get tip-tilt information. We show a  ``best-case'' scenario for LTAO, where the tip-tilt NGS is on-axis. To show that the tip-tilt determination is working correctly, we also simulate an LGS tomographic system where LGS tip-tilt information is removed but no NGS is used for tip-tilt correction. \changed{In all simulated AO modes the lasers are launched from the centre of the telescope aperture, behind the secondary obscuration.}

The code used to perform these simulations has been written solely in the Python programming language and incorporates full physical optical propagation of LGS as they pass up through the atmosphere to form a guide star. This is necessary to accurately estimate the LGS up-link path as the Fresnel number for a typical LGS beam is $\approx 1$ , meaning diffraction cannot be ignored \citep{holzlohner2008}. A geometric ray tracing method is used to calculate the wavefront measured on the WFS from turbulence encountered as light passes down through the atmosphere. The PSF formed by the LGS up-link is then convolved with each sub-aperture PSF to give realistic simulation of LGS up-link turbulence. Focal anisoplanatism (cone effect) is included when propagating light down from an LGS. These simulations do not include either read noise or photon shot noise. Simulation parameters are shown in Table \ref{fig:simParams}. The code, ``Soapy'', is available publicly and is free for use\footnote{https://github.com/andrewpaulreeves/soapy}.

\changed{The phase screens used in these simulation have been made significantly larger than the size of the simulated telescope aperture. This mitigates the periodic nature of some phase screen generation algorithms and ensures that there is sufficient power in low order modes at the spatial scale of the telescope aperture \citep{schmidt2010}.}

\begin{table}
  \label{fig:simParams}
  \caption{Parameters used in the simulation of LGS up-link tip-tilt retrieval} 
  \begin{tabular}{ | p{6cm} || p{1.5cm} | }
    \toprule
	Parameter					&		Value	\\
    \midrule
	Telescope primary diameter (m)	&		8\\
    Central obscuration diameter (m) &       1.1\\
	Phase points in pupil		&		256 $\times$ 256 \\
	Total phase screen size (m)			&		128 $\times$ 128 \\
	Radius of off-axis LGS asterism (arc-seconds)	&		10 \\
	WFS sub-apertures			&		16$\times$16 \\

	DM Actuators				&		17$\times$17\\
	
	Science field wavelength ($\mu$m)	&		1.65 \\
	Science integration time (s)	&		60 \\
	Loop frame rate	(Hz) 		&		400 \\
    \bottomrule
  \end{tabular}
\end{table}

\subsection{Simulated covariance matrices}

\begin{figure}
 \centering
   \includegraphics[width=.5\textwidth, trim={0.5cm 0.5cm 1.cm 0.5cm },clip=true]{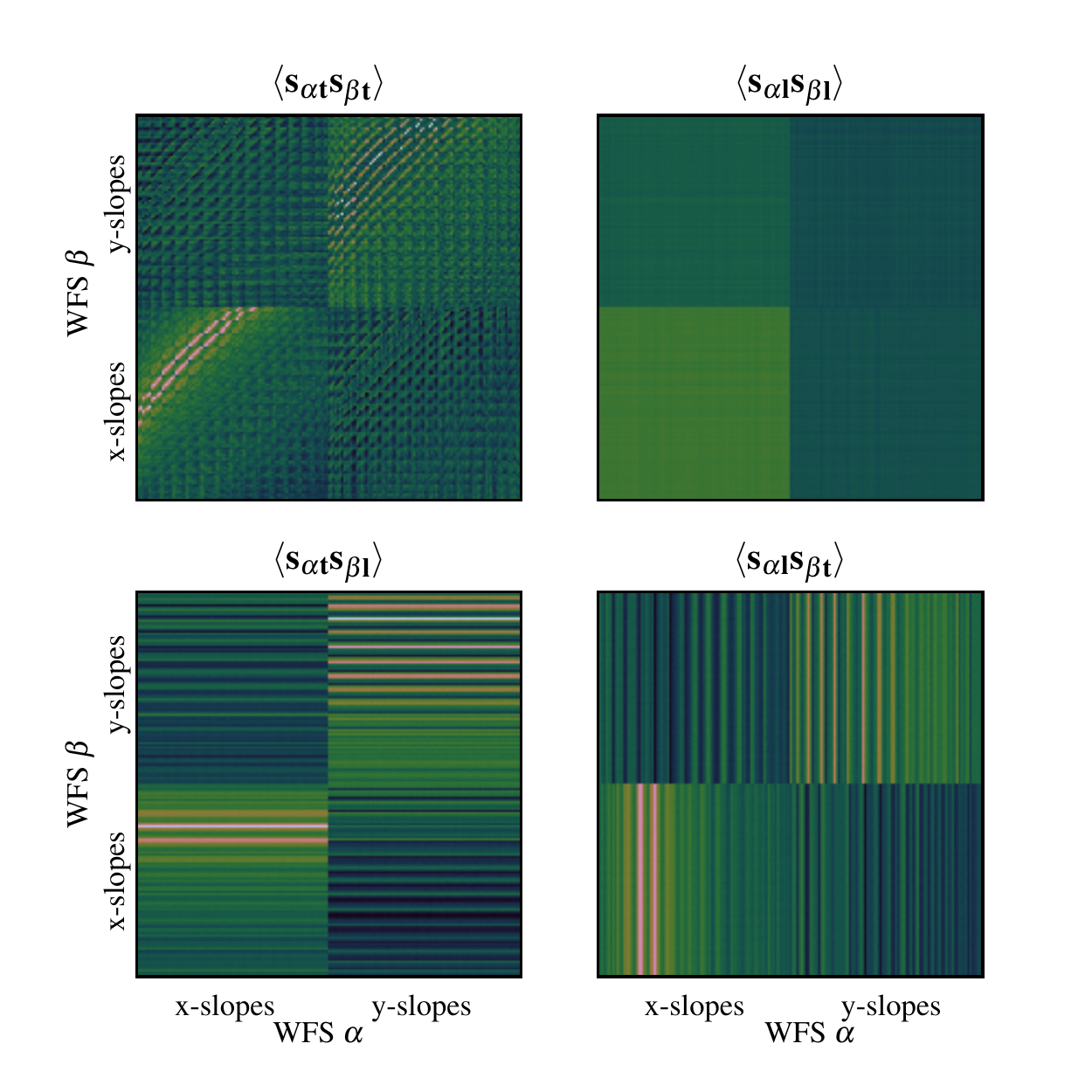}
 \caption{A simulated LGS up-link covariance matrix between two LGS WFSs observing in different directions deconstructed into its constituent terms. The top-left image shows the covariance of down-link turbulence. The bottom two images show the covariance of up-link turbulence with down-link turbulence resulting in characteristic ``stripes''. The top-right image shows covariance between up-link turbulence, which is constant for each x-y combination.}
 \label{fig:covmat}
\end{figure}

The creation of the tomographic LGS tip-tilt reconstructor depends upon the covariance matrices between the various WFSs, the form of which was derived in equation (\ref{eq:covmat}). For the results presented in this paper we simulate the covariance matrices by recording many open loop AO system frames on uncorrelated turbulence phase screens, until the covariance matrices converge to the theoretical case. This process must be repeated for every atmospheric and AO configuration simulated. 

An example simulated covariance matrix between two LGS WFSs with up-link included, with a single turbulence layer at an altitude of $\mathrm{14}$km, is shown in Figure \ref{fig:covmat}. The covariance matrix has been deconstructed into its constituent terms by simultaneous simulation of LGS with and without the effects of LGS up-link tip-tilt observing in the same direction. The final covariance matrix used to create the tip-tilt LGS tomographic reconstructor is the sum of these terms. 

From Figure \ref{fig:covmat}, it is possible to see that the terms match our qualitative predictions. The first term, $\langle s_{\alpha t} s_{\beta t} \rangle$, (top-left) is identical to that caused by down-link turbulence and hence looks similar to those used for conventional LA LTAO, though with down-link tip-tilt included. The final term, $\langle s_{\alpha l} s_{\beta l} \rangle$, (top-right) is the result of covariance between the path of the two lasers and is hence constant for each x-y pair. The middle terms, $\langle s_{\alpha t} s_{\beta l}\rangle$ and $\langle s_{\alpha l} s_{\beta t}  \rangle$, (bottom left and right) are dominated by the strong covariance between up-link tip-tilt on one WFS and a small number of sub-apertures on the other WFS which observe its up-link path. Hence they are seen as horizontal and vertical ``stripes'' of high slope covariance.

\subsection{Performance versus turbulence altitude}

\begin{figure}
    \centering
        \includegraphics[width=0.5\textwidth]{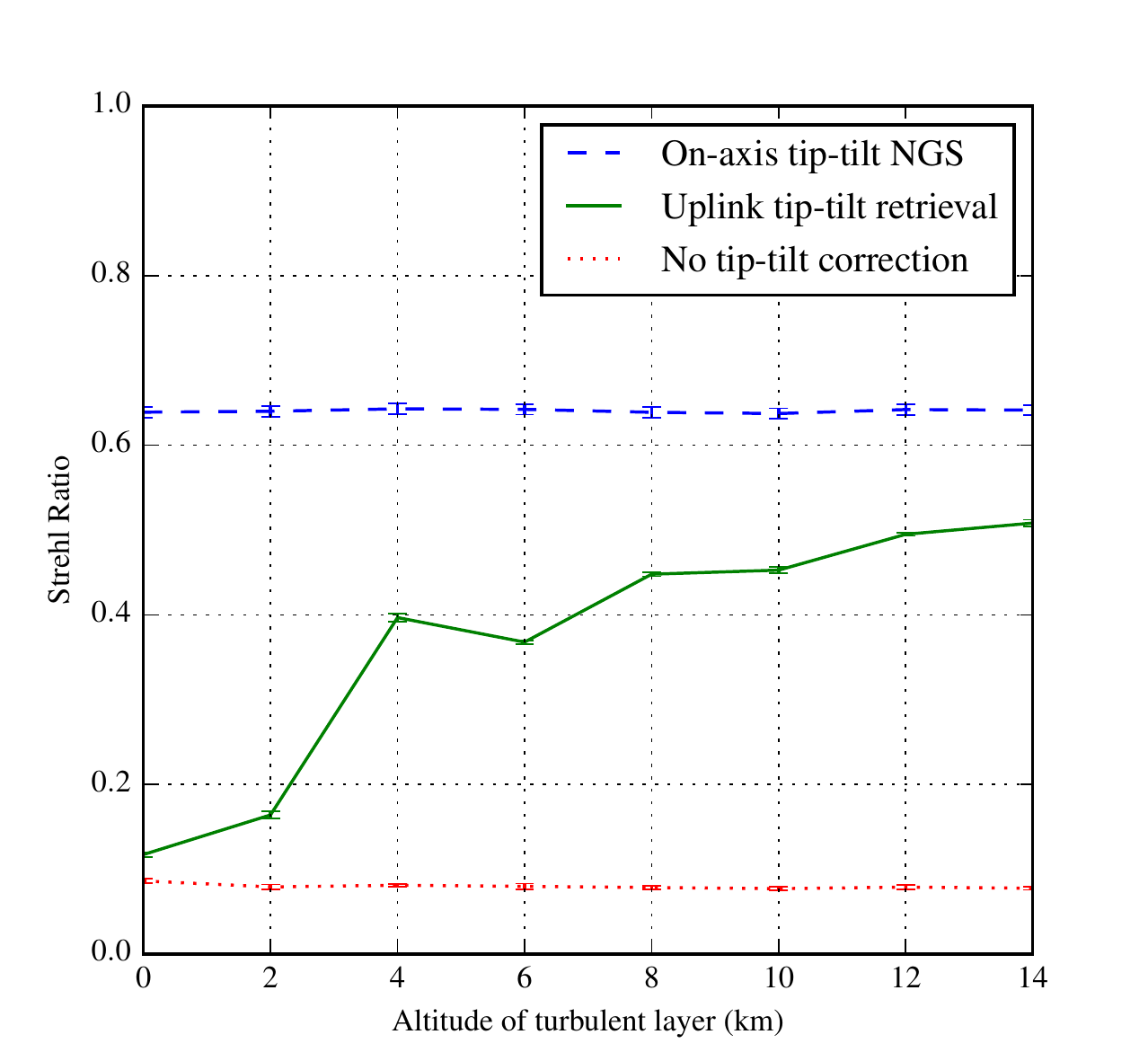}
    \caption[Strehl ratio of tip-tilt retrieval versus altitude of single atmospheric turbulence layer]{The Strehl ratio of tomographic LGS AO with a NGS tip-tilt sensor (dashed), the tip-tilt retrieval method (solid) and no tip-tilt correction (dotted) versus the altitude of a single turbulence layer. The LGS up-link tip-tilt determination performance increases sharply as the layer altitude increases.}
    \label{fig:strehlVsHeight}
\end{figure}    

In § \ref{sec:lgsTT} and § \ref{sec:LA} it was predicted that the method of including LGS up-link tip-tilt would be more effective for turbulence at higher altitudes. To further investigate this hypothesis, AO correction performance versus the altitude of a single turbulence layer is simulated, with results shown in Figure \ref{fig:strehlVsHeight}. 

In line with our predictions, the results show low performance of LGS up-link tip-tilt determination when low layer turbulence is present. More promisingly, they also show that the methods works well to correct high layer turbulence, where the correction may even approach that of LTAO using an on-axis NGS for tip-tilt correction. Discontinuities in the curve are due to the up-link beam overlapping different numbers of sub-apertures at different altitudes. 

\subsection{Performance with multi-layer turbulence profile}
\begin{figure}
    \centering
    \includegraphics[width=0.5\textwidth]{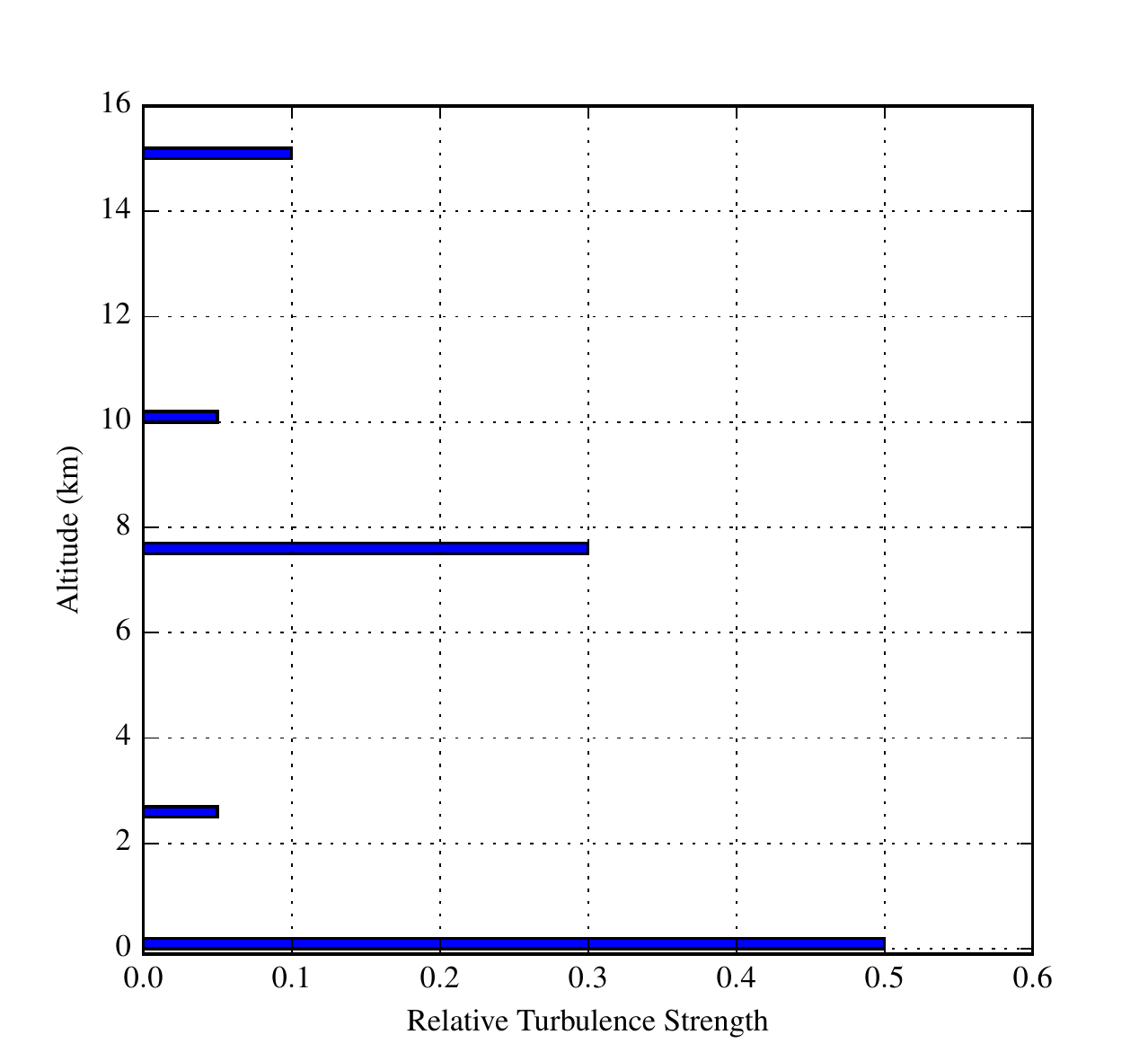}
    \caption{The atmospheric turbulence profile used in the simulations. It contains high layers of significant strength, which will require a tomographic system to correct. The profile also includes a strong ground layer, which we do not expect to be well corrected by the LGS tip-tilt determination.}
    \label{fig:profile}
\end{figure}

To give an impression of how the algorithm may perform with no NGS under more realistic atmospheric turbulence conditions, we perform simulations with the profile shown in Figure \ref{fig:profile}. This contains a strong ground layer, which is unlikely to be well corrected by the up-link tip-tilt determination, as well as significant higher layers, for which the benefit of including the tip-tilt correction may become apparent.

\begin{figure}
  \centering
    \includegraphics[width=.5\textwidth]{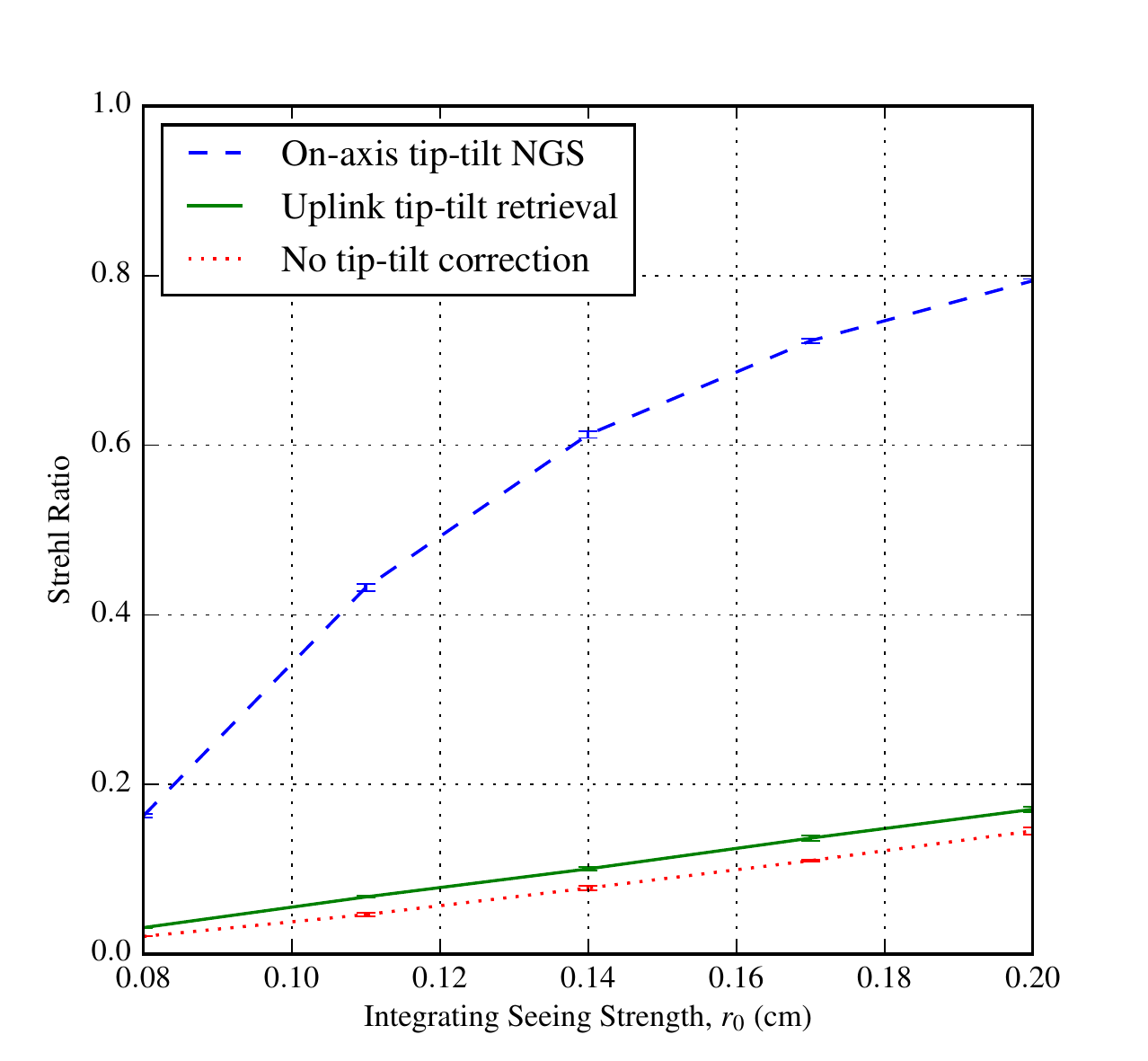}
  \caption{Performance of LTAO with an on-axis tip-tilt NGS (dashed), with no NGS and the tip-tilt retrieval (solid) and with no tip-tilt correction (dotted). The tip-tilt retrieval method performs slightly better than with no tip-tilt correction, but the presence of ground layer turbulence has degraded performance significantly.}
  \label{fig:plots_strehlvsR0}
\end{figure}

Figure \ref{fig:plots_strehlvsR0} shows the performance of LTAO with an on-axis NGS tip-tilt, LTAO with tip-tilt retrieval and LTAO with no tip-tilt correction versus increasing seeing strength. With a strong ground layer present the method is clearly not as effective as when there is only high layer turbulence. It does still provide a small improvement over correction with no tip-tilt correction. 

\begin{figure}
  \centering
    \includegraphics[width=.5\textwidth]{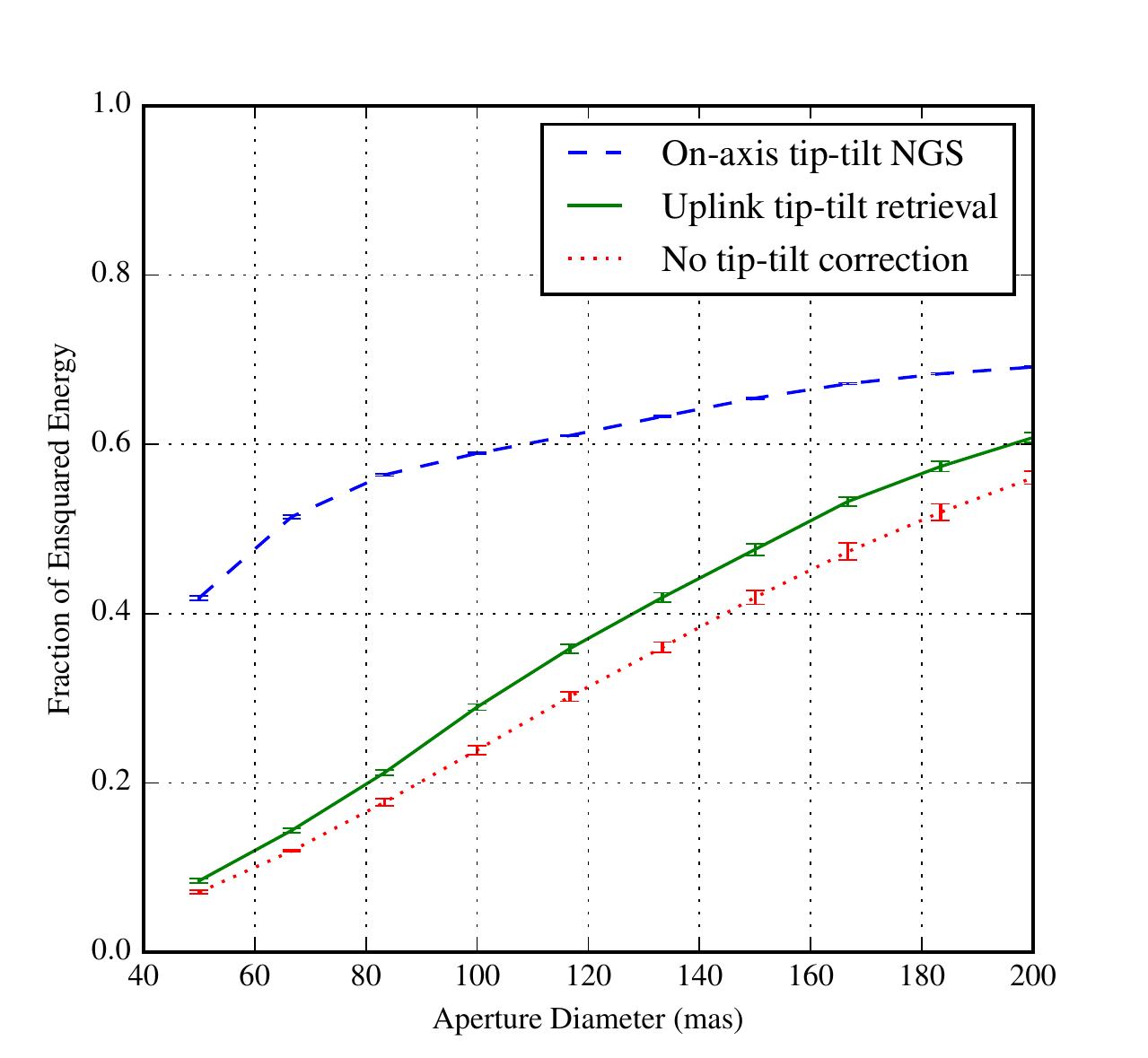}
  \caption{Ensquared energy versus aperture size of LTAO with an on-axis tip-tilt NGS(dashed), with no NGS and the tip-tilt retrieval (solid) and with no tip-tilt correction(dotted). Though never approaching that of LTAO with an on-axis NGS, the tip-tilt retrieval improves throughput over the case when no tip-tilt correction is performed.}
  \label{fig:plots_AOvsEE_r0-14cm}
\end{figure}

To further investigate how this small improvement could aid spectrographic instruments, the Ensquared Energy into a potential spectrograph spaxel is plotted against the size of the spaxel in Figure \ref{fig:plots_AOvsEE_r0-14cm}. For these results the Fried parameter, $\mathrm{r_0}$, is $\mathrm{14}$ cm. As previously discussed, \citet{davies2008} have explored how LGS AO with no tip-tilt correction may be useful for some science cases due to the increased throughput. Our method can provide higher throughput, still without the use of any NGSs. For instance, a spectrograph with spaxel size of 100 mas will receive around 5\% more light per spaxel for our simulate conditions. 

\section{Discussion}
\label{sec:discussion}

\subsection{Improvement of LGS only AO}
The method presented in this paper has been shown to correct well for high layer turbulence above a tomographic LGS AO system. As was predicted in the derivation of the algorithm, it does not correct well for low layer turbulence. At most major observing sites the ground layer of turbulence is a significant contributor to overall seeing strength \citep{chun2009, osborn2010, garcialorenzo2011}, especially when combined with other effects such as seeing caused by the telescope structure itself \citep{berdja2013, shepherd2014}. \changed{It is also possible that a turbulence profile containing more than five layers may further reduce performance further. A profile with only five layers was chosen due to the computational load of performing physical light propagation between each layer. It is intended to optimise the code to allow simulations with a higher number of turbulence layers to be performed for future works.} With this in mind, the up-link tip-tilt method alone is unlikely to provide AO correction approaching that of LTAO with NGS tip-tilt guide stars. 

However, for some science cases which require low-order correction in a section of the sky sparsely populated by suitable tip-tilt NGS, accounting for up-link turbulence may still be of some use.  We have shown that it \changed{may} provide a modest improvement in system throughput. It can be implemented without great hardware modification, and only requires a change in tomographic reconstructor in the real-time controller.

\subsection{Combination with ground layer NGS tip-tilt correction}
The greatest potential impact of this method of accounting for LGS up-link turbulence is in combination with a number of far off-axis tip-tilt NGS. With the LGS tip-tilt retrieval, high layer turbulence can be corrected for, and the LGS can still correct for high order ground layer turbulence. This leaves only turbulence near the ground for which the AO system requires tip-tilt information. The ground layer is common to all directions, so can consequently been corrected by a number of very far off-axis NGS. This can greatly increase the sky-coverage of LTAO systems.

On the other hand, performance could be improved for existing LTAO configurations when using the current furthest off-axis tip-tilt NGS. In this case much of the wavefront error is from high altitude tip-tilt modes which are not well sampled by the far off-axis tip-tilt NGS. If the LGS tip-tilt up-link determination is implemented, performance may be significantly increased. 

In future work we will continue to further investigate the implementation of an AO configuration where ground layer tip-tilt is corrected using far off-axis NGS tip-tilt references. We will also examine the potential performance and sky coverage gains of existing LTAO with far off-axis tip-tilt.

\section{Conclusions}
\label{concs}

We have demonstrated theoretically and in simulation the viability of a tomographic LGS reconstructor which determines tip-tilt information by accounting for the up-link path through atmospheric turbulence of each LGS. This is proven possible geometrically and implemented using a LA based technique to utilise information based on correlations of adjacent sub-apertures. 

The algorithm shows good performance when correcting for high layer turbulence, close to that of LTAO with an on-axis tip-tilt NGS. The performance when low altitude turbulence is present is much degraded, though still an improvement over having no tip-tilt correction. 

Though the method may be of some use as an LGS only AO reconstructor for low spatial order science cases, we mainly envisage it augmenting LTAO and MOAO by allowing further off-axis tip-tilt NGSs and hence improving AO corrected sky coverage. It also allows for greater LTAO performance with existing tip-tilt NGS seperations. In future works we will expand on these themes and quantify the available improvements.

One major conclusion from this work is that LGS up-link turbulence does not have to be simply ignored and discarded as is currently the case. It is not irretrievably entangled with the down-link turbulence, and as such can provide useful information about the atmosphere above the telescope.

\section*{Acknowledgments}
APR gratefully acknowledges support from the Science and Technology Facilities Council (STFC) in the form of a Student Enhancement Program (STEP) award (grant code: ST/J501013/1). TJM, RMM, NAB and JO also acknowledge financial support from STFC (grant code: ST/L00075X/1).

The data used for the results in this publication is available on request to the author.




\bibliographystyle{mnras}
\bibliography{bibliography} 




\appendix
\section{Formulating the relationship between a sub-aperture measurement and up-link LGS jitter.}

\label{append:Interaction}
\begin{figure}
	\centering
	\includegraphics[width=0.5\textwidth]{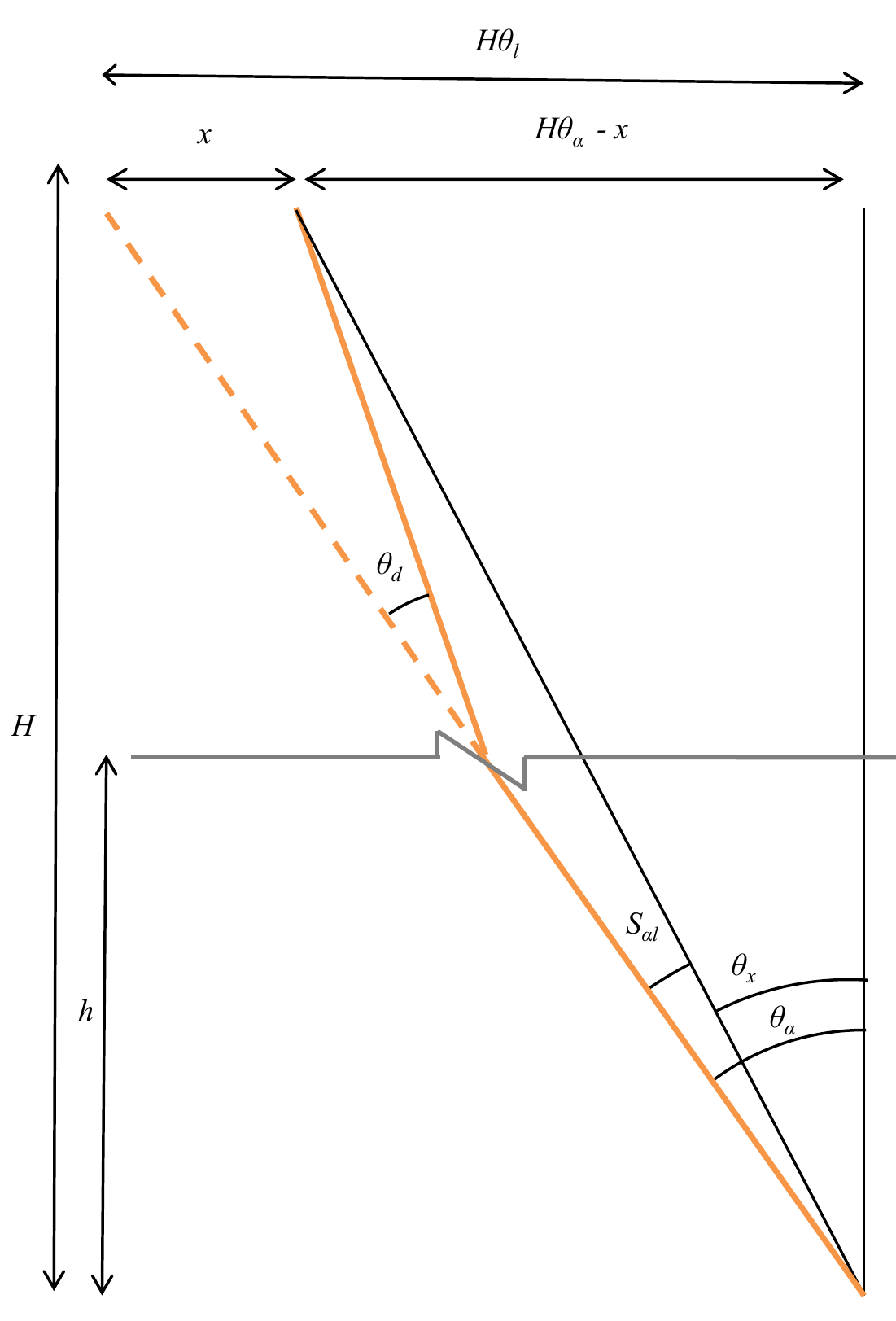}
	\caption{The displacement on LGS $\alpha$ caused by a turbulence encountered on up-link.}
	\label{fig:t_matrix}
\end{figure}
To find the relationship between a sub-aperture measurement on WFS $\beta$, $s_{\beta t}$ and the measured up-link on LGS $\alpha$, $s_{\alpha l}$ we consider Figure \ref{fig:t_matrix}, which shows the effect of a small tilt on LGS $\alpha$ up-link.
\begin{equation}
	s_{\alpha l} = \theta_{\alpha} - \theta_{x},
\end{equation}
where $\theta_{\alpha}$ is the launch angle of LGS $\alpha$ and $\theta_x$ is dependent on the tip-tilt LGS $\alpha$ passes through at height $\mathrm{h}$.
\begin{equation}
	\theta_x = \frac{H \theta_l - x}{H}
	\end{equation}
and
\begin{equation}
	x = \theta_d (H-h),
	\end{equation}
	so 
\begin{align}
	s_{\alpha l} 	&= \theta_l - \frac{H\theta_l - \theta_d(H-h)}{H} \nonumber \\					&= \theta_d \frac{H-h}{H}.
	\end{align}

\section{Calculating resolved tomographic bin heights}
	\label{append:binHeight}
		Only a finite number of LGS up-link offsets at a turbulent layer can be predicted by the above method. This is a result of finite resolution of the WFS being used. For a centre launched tomographic LGS, only half the number of WFS sub-apertures view the path of another LGS, hence only half can be used to predict up-link LGS motion. 
	
		Turbulence which affects the path of an LGS can only be measured in vertical `bins' where the measurement is the sum of the turbulence in the sub-aperture FOV within the vertical bin. The heights of these `bins' can be calculated by considering the geometry of the system, again illustrated in Figure \ref{fig:overlapping_LGS}. $\theta_\alpha$, $\theta_\beta$ are the launch angles for LGS $\alpha$ and $\beta$ respectively, $H$ is the height of the LGS constellation, $D_s$ is the displacement for the centre of sub-aperture $n$ and the LGS launch position with is assumed to be the centre of the telescope pupil, $D$ is the diameter of the telescope pupil.
	
For small angles
		\begin{equation}
			h = \frac{D_s}{\theta_s + \theta_{\alpha}}.
			\end{equation}
$\theta_s$ can be obtained by considering the displacement on the ground between LGS position and the centre of the sub-aperture, $D_s+H\theta_{\beta}$.
		\begin{equation}
			\theta_s = \frac{D_s+H\theta_{\beta}}{H}.
			\end{equation}
$D_s$ is dependent on the sub-aperture of interest, $n$, where the position of the centre of a sub-aperture is $(n+0.5)d$, and $d$ is the diameter of sub-aperture.
		\begin{equation}
			D_s = \frac{D}{2} - (\mathrm{n}+0.5)\mathrm{d}
		\end{equation}
	
	Finally, an expression for the height of the centre of the resolved height bin is, $h$, can be obtained.
	\begin{align} 
		h &= \frac{D_s}{\frac{D_s+H\theta_\beta}{H}+\theta_{\alpha}} \\
		h &= \frac{HD_s}{D_s+H(\theta_{\alpha}+\theta_{\beta})} \\
		h_n &= \frac{H(\frac{D}{2} - (n+0.5)d)}{\frac{D}{2} - (n+0.5)d+H(\theta_{\alpha}+\theta_{\beta})}
	\end{align}
	

\bsp	
\label{lastpage}
\end{document}